\documentstyle[aps]{revtex}

\input epsf         
\epsfverbosetrue    
\def\postscript#1{\begin{center}\leavevmode
\hbox{\epsfxsize=0.95\columnwidth\epsfbox{#1}}\end{center}}

\begin{document}

\twocolumn[\hsize\textwidth\columnwidth\hsize\csname@twocolumnfalse%
\endcsname

\draft

\title{C-axis Transport in Bilayer Cuprates and
Relation to Pseudogap}

\author{W. C. Wu$^a$, W. A. Atkinson$^b$, and J. P. Carbotte$^a$}
\address{$^a$Department of Physics and Astronomy, McMaster University,
Hamilton, Ontario, Canada L8S 4M1}
\address{$^b$Department of Physics, Indiana University,
Swain Hall W 117, Bloomington, IN 47405}
 
\date{\today}

\maketitle

\begin{abstract}
We consider the effect of interband transitions on
the $c$-axis conductivity, DC resistivity, and thermal conductivity
in a plane-chain bilayer model of a cuprate. 
The relation between the $c$-axis resistivity and thermal conductivity
is governed by the Wiedemann-Franz law. When the perpendicular hopping 
matrix element between chain and plane ($t_\perp$) is small, 
the $c$-axis DC resistivity shows a 
characteristic upturn as the temperature is lowered and the infrared 
conductivity develops a pseudogap. As
$t_\perp$ is increased, intraband transitions start to dominate and a more
conventional response is obtained. Analytical results 
for a simple plane-plane bilayer are also given, including the
frequency sum rule of the optical conductivity.
\end{abstract}

]

\vskip 0.1 true in
\narrowtext

\newpage

\section{Introduction}

In a previous paper \cite{AC96-1}, we derived expressions for,
and computed, the normal-state optical conductivity for a system of two layers
per unit cell coupled through a hopping matrix element $t_\perp$ which 
introduces the third dimension. In addition to the usual intraband terms, 
which give a Drude-like response, interband transitions
are now possible and these have quite a different temperature 
and frequency dependence.
It was found that for sufficiently small values of $t_\perp$, the interband 
terms, which are of order $t_\perp^2$, dominate over
the intraband contribution which goes like $t_\perp^4$ for the $c$-axis.
In this limit, the $c$-axis response can be
very different from Drude. In general, the
interband transitions occur at energies which
extend over the entire band width and structure in the
$c$-axis conductivity can be
directly related to certain features of the band structure.
For the particular case of a CuO$_2$ tetragonal plane and a set of
CuO chains \cite{BDLH,Zhang,Bonn}, it was found that
the normal-state $c$-axis response can develop
a gap or pseudogap at low $\omega$ depending on the details of the two
Fermi surfaces involved.

Estimated values of $t_\perp$ have been provided by Zha, Cooper and Pines 
\cite{ZCP}. They give $t_\perp\sim 3.0{\rm meV}$ for
underdoped YBCO$_{6.68}$ and 30 to 40{\rm meV} for
optimally doped YBCO$_{6.93}$. These estimates are
much larger than for the more
anisotropic compounds such as Bi$_2$Sr$_2$CaCu$_2$O$_8$ where
$t_\perp$ is perhaps as small as $0.1{\rm meV}$. It is
clear from these estimates that underdoped
YBCO could be in the interband dominated
regime and show a pseudogap response while
optimally doped and overdoped YBCO would
show a more usual Drude-like $c$-axis response as in this case,
intraband transitions would dominate over
interband.

In this paper, we wish to consider the $c$-axis
transport and study the effect of interband transitions on its
temperature dependence. For fully oxygenated YBCO, the
resistivity $\rho(T)$ in the $c$-axis
\cite{Friedmann90,Welp,Iye,GLT,Takei,Iye88,BWO,TMTU} 
is found to exhibit metallic behavior
and show a nearly linear temperature dependence
which follows that in the $a$-$b$ plane but, of course, has
a very different magnitude with $\rho_c/\rho_{ab}$ of order 30 to 70.
On the other hand, for deoxygenated  underdoped
samples corresponding to small $t_\perp$, the situation is very different and 
$\rho_c(T)$ shows a semiconductivity-like behavior 
\cite{Iye,Takei,Iye88,BWO,TMTU}
at low temperatures ($d\rho_c/dT<0$), while
at high temperatures, the linear behavior is retained.
The ratio $\rho_c/\rho_{ab}$ is however
very much larger than that observed
in the fully oxygenated case with  $\rho_c/\rho_{ab}$
at $T=100K$ of order $10^3$. Estimates of $c$-axis mean free paths based 
on such large values of $\rho_c$ lead to values that are of the 
order and even less than the $c$-axis spacing between planes.
This realization has led some to suggest that conventional 
three dimensional Bloch transport breaks down and that
$c$-axis transport proceeds through an incoherent 
mechanism due to confinement coming from the highly
two dimensional nature of the electronic
states. Here we consider that the interband transitions begin to dominate 
when $t_\perp$ is small. This effect has its origin in the fact that YBCO 
has several atoms per unit cell. In the real situation, there are two
copper-oxygen planes and one set of chains. For simplicity, here
we will consider only a plane-chain model
which is believed to capture some of the qualitative features
of the interband transitions.

In Sec.~II, we establish the Kubo
formula for the AC infrared conductivity in the case of a plane-chain 
model coupled through a transverse hopping matrix element $t_\perp$ in
tight binding. The usual formula becomes a trace over
the two bands and the electromagnetic 
vertices are now a $2\times 2$ matrix
with off-diagonal elements related to the derivatives
of the Hamiltonian matrix. The resulting formula easily
separates into interband and intraband parts. 
In Sec.~III, analytic results for the 
infrared conductivity and the DC resistivity 
are presented for the case of infinite bands
with minimum energy difference between them of $\Delta$. One band
is taken to have a one-dimensional character and the other two-dimensional.  
The role of the pseudogap $\Delta$ in these quantities is made clear.
In Sec.~IV, the thermal conductivity is considered and the Wiedemann-Franz law 
is established for the $c$-axis including interband as well as
intraband contributions.  In Sec.~V, we treat the case 
of a plane-plane bilayer which is important because simpler analytic 
results can be obtained including for the conductivity sum rule.
In Sec.~VI, tight-binding bands are introduced and numerical results
presented for this more complicated case which does not yield to any 
analytic analysis but includes such  effects as the Van Hove 
singularity in the electronic density of states. Nevertheless the 
results show no qualitative difference with those presented in previous
sections based on simplified infinite bands.
A short conclusion can be found in Sec.~VII.

\section{Formalism}

We start with the Hamiltonian for a plane-chain bilayer coupled through a 
transverse matrix element $t_\perp$. In a matrix 
notation, we start with creation 
$C^\dagger_{\bf k}$ and annihilation $C_{\bf k}$ operators

\begin{eqnarray}
C^\dagger_{\bf k}=\left(c^\dagger_{1{\bf k}},
                    c^\dagger_{2{\bf k}}\right) ~~~~;~~~~~~~
C_{\bf k}=\pmatrix{c_{1{\bf k}}\cr
                    c_{2{\bf k}}\cr},
\label{eq:c}
\end{eqnarray}
with $c^\dagger_{1{\bf k}}(c^\dagger_{2{\bf k}})$ creating an electron
in the state of momentum ${\bf k}$ in layer 1 (2), respectively.
The Hamiltonian operator $H_0$ is
\begin{equation}
H_0=\sum_{\bf k}C^\dagger_{\bf k}h({\bf k})C_{\bf k},
\label{eq:H0}
\end{equation}
where the Hamiltonian matrix

\begin{equation}
h({\bf k})= \pmatrix{\xi_1({\bf k})&t(k_z)\cr
                t(k_z)&\xi_2({\bf k})\cr},
\label{eq:h}
\end{equation}
with
\begin{mathletters}
\label{eq:simpleband}
\begin{eqnarray}
\label{eq:simpleband.xi1}
\xi_1({\bf k})&=&{\hbar^2\over 2m}(k_x^2 +k_y^2)-\mu+\Delta,\\
\xi_2({\bf k})&=&{\hbar^2 \over 2m}k_y^2-\mu,
\label{eq:simpleband.xi2}
\end{eqnarray}
\end{mathletters}
for the plane and chain respectively and 
\begin{eqnarray}
t(k_z)=-2t_\perp \cos\left({k_z d\over 2}\right)
\label{eq:t}
\end{eqnarray}
with $d$ the distance between chain
and plane in the $z$-direction and $t_\perp$ the coupling strength. 
In (\ref{eq:simpleband}), $\mu$ is the chemical potential,
$\Delta$ is the minimum energy difference between the 
two bands and is responsible for the pseudogap in the $c$-axis response, 
$\hbar$ is the Planck's 
constant over $2\pi$ and $m$ is the bare electron mass.

When a vector potential ${\bf A}({\bf q})$ is applied
to the system, the Hamiltonian changes to \cite{Mahan,AC95}

\begin{eqnarray}
H=H_0-{e\over 2mc}{\bf p}\cdot {\bf A}
\label{eq:H0.A}
\end{eqnarray}
to the lowest order in ${\bf A}$, with the vector

\begin{equation}
{\bf p}={m\over \hbar}
\sum_{\bf k}C^\dagger_{\bf k}{\partial h({\bf k})\over
\partial {\bf k}} C_{\bf k}.
\label{eq:p}
\end{equation}
Using the Kubo formula for the current-current correlation function,
this leads directly to a real-part conductivity in the $c$-axis of the form
(${\rm Re}\sigma_{zz}(\omega)\equiv\sigma_c(\omega)$)

\begin{eqnarray}
&&\sigma_c(\omega)=e^2{1\over \Omega}
\sum_{\bf k}\int_{-\infty}^{\infty}{dx\over 2\pi}~
{f(x)-f(x+\hbar\omega)\over \omega}\nonumber\\
&&\times {\rm Tr} [A({\bf k},x)\gamma_z({\bf k,k})
A({\bf k},x+\hbar\omega)\gamma_z({\bf k,k})],
\label{eq:conductivity}
\end{eqnarray}
where ``Tr'' is a trace, $A({\bf k},\omega)$ 
is the spectral function matrix, and
$\gamma_z$ is the associated $c$-axis electromagnetic
vertex function matrix given by

\begin{eqnarray}
\gamma_z={1\over \hbar}{\partial h({\bf k})\over \partial k_z}.
\label{eq:gammaz.def}
\end{eqnarray}

The ``trace'' operator in (\ref{eq:conductivity}) allows one to work in 
any frame which is convenient. We will choose a frame in which the 
Green's function (and hence the spectral function) matrix is diagonal. 
Consequently, one obtains (following Ref.~\cite{AC96-1})

\begin{eqnarray}
&&\sigma_{c}(\omega)=e^2{1\over \Omega}
\sum_{\bf k}\int_{-\infty}^{\infty}{dx\over 2\pi}~
{f(x)-f(x+\hbar\omega)\over \omega}
\label{eq:rhoc} \\
&&\times\Biggl[\Bigl(A_{11}A_{11}^\prime+ A_{22}A_{22}^\prime \Bigr)
\gamma^2_{11}+
\Bigl(A_{11}A_{22}^\prime+ A_{22}A_{11}^\prime \Bigr)
\gamma^2_{12} \Biggr],\nonumber
\end{eqnarray}
where $A_{ii}\equiv A_{ii}({\bf k},x)$ and 
$A_{ii}^\prime\equiv A_{ii}({\bf k},x+\hbar\omega)$
and the vertices

\begin{eqnarray}
\gamma_{11}&=&
v_\perp\left({2t\over\epsilon_+ -\epsilon_-}\right)\nonumber\\
\gamma_{12}&=&
v_\perp\left({\xi_1 -\xi_2\over\epsilon_+ -\epsilon_-}\right),
\label{eq:vertices}
\end{eqnarray}
with $v_\perp\equiv 1/\hbar\partial t/\partial k_z=
t_\perp d/\hbar \sin(k_z d/2)$.
More explicitly, in the normal state, the spectral functions

\begin{equation}
A_{ii}({\bf k},x)={2\Gamma_i\over (x-\epsilon_i)^2+\Gamma_i^2},
\label{eq:spectral}
\end{equation}
where $\epsilon_1\equiv \epsilon_+$, $\epsilon_2\equiv \epsilon_-$,
and $\Gamma_i$ is the total scattering rate in band $i$.
A realistic model for the
high-$T_c$ oxides is that the scattering rate is dominated by 
the inelastic scattering with $\Gamma$ linear in temperature $T$.
The energies 

\begin{equation}
\epsilon_\pm={\xi_1+\xi_2\over 2}\pm
\sqrt{\left({\xi_1-\xi_2\over 2}\right)^2+t^2}
\label{eq:spectrum2}
\end{equation}
are the two renormalized bands. 

It is clear from the form of Eq.~(\ref{eq:rhoc}) that with the
initial and final states in the same band, the first 
term corresponds to the {\em intraband} contribution and the second 
term corresponds to the {\em interband} contribution
with the initial and final states in different bands.
From the form of the vertices given in (\ref{eq:vertices}), one 
sees immediately 
that the interband contribution ($\sim t_\perp^2$) dominates
the intraband contribution ($\sim t_\perp^4$) in the small-$t_\perp$ limit.
For simplicity, we will assume the scattering rate
$\Gamma_1=\Gamma_2\equiv \Gamma$ in all our calculations.

\section{Conductivity and DC Resistivity}

Substituting (\ref{eq:simpleband}), 
(\ref{eq:vertices}), and (\ref{eq:spectral}) into (\ref{eq:rhoc}),
the intraband contribution for the $c$-axis conductivity
is explicitly given by  
 
\begin{eqnarray}
&&~~\sigma_{c}^{\rm intra}(\omega)=e^2{1\over \Omega}
\sum_{\bf k}v_\perp^2\left({2t\over \epsilon_+ - \epsilon_-}\right)^2
\int_{-\infty}^{\infty}{dx\over 2\pi} \nonumber\\
&&{f(x)-f(x+\hbar\omega)\over \omega}
\Biggl[{2\Gamma\over (x-\epsilon_+)^2+\Gamma^2}\
{2\Gamma\over (x+\hbar\omega-\epsilon_+)^2+\Gamma^2}\nonumber\\
&&~~~~+{2\Gamma\over (x-\epsilon_-)^2+\Gamma^2}\
{2\Gamma\over (x+\hbar\omega-\epsilon_-)^2+\Gamma^2} \Biggr].
\label{eq:rhoc.intra}
\end{eqnarray}
To reduce (\ref{eq:rhoc.intra}) further, we first replace the momentum 
${\bf k}$-sum by an integral
 
\begin{equation}
{1\over \Omega} \sum_{\bf k}\rightarrow
\int_{-{\pi\over d}}^{\pi\over d}{dk_z\over 2\pi}
\int_{-\infty}^{\infty}N(0)d\epsilon,
\label{eq:sum.approxi.1}
\end{equation}
where $\epsilon$ can be either $\epsilon_+$ or $\epsilon_-$ and
a constant ``2D'' density of states per spin at Fermi level 
$N(0)\equiv m/2\pi\hbar^2$ is 
assumed (in the small-$t_\perp$ limit).
Secondly, we replace the factor $\epsilon_+ - \epsilon_-$ in 
(\ref{eq:rhoc.intra})
by the (constant) band gap $\Delta$. This approximation is quite good and
has only a small effect on the final results.  
As a result of the above simplification, 
the $k_z$ dependence in the integrand comes 
in only through $t(k_z)$ and $v_\perp(k_z)$
which can be easily carried out to get
$\int_{-{\pi\over d}}^{\pi\over d}
{dk_z\over 2\pi}v_\perp^2 t^2=t_\perp^4 d/ 2\hbar^2$.
Consequently, (\ref{eq:rhoc.intra}) reduces to
 
\begin{eqnarray}
&&\sigma_{c}^{\rm intra}(\omega)=
{8e^2t_\perp^2N(0)d\over\pi\hbar^2}\left({t_\perp\over\Delta}\right)^2
\int_{-\infty}^{\infty}{d\epsilon} \int_{-\infty}^{\infty}dx
\label{eq:rhoc.intra.1}\\
&&{f(x)-f(x+\hbar\omega)\over \omega}
{\Gamma\over (x-\epsilon)^2+\Gamma^2}\
{\Gamma\over (x+\hbar\omega-\epsilon)^2+\Gamma^2},\nonumber
\end{eqnarray}
where the two bands give the {\em same} contributions for the intraband part
of the $c$-axis conductivity.  Furthermore, with the aid of
 
\begin{eqnarray}
&&\displaystyle \int_{-\infty}^{\infty}{d\epsilon}
{\Gamma\over (x-\epsilon)^2+\Gamma^2}\
{\Gamma\over (x+\hbar\omega-\epsilon)^2+\Gamma^2}\nonumber\\
&&~~~~~~~~=\pi{2\Gamma\over (\hbar\omega)^2+4\Gamma^2}
\label{eq:int1}
\end{eqnarray}
and

\begin{eqnarray}
\int_{-\infty}^{\infty}dx~{f(x)-f(x+\hbar\omega)\over \omega}
=\hbar, 
\label{eq:int2}
\end{eqnarray}
one immediately obtains the final result
 
\begin{equation}
\sigma_{c}^{\rm intra}(\omega)=
{8e^2t_\perp^2N(0)d\over \hbar}\left({t_\perp\over\Delta}\right)^2
{2\Gamma\over (\hbar\omega)^2+4\Gamma^2}
\label{eq:Drude}
\end{equation}
which is nothing but the ``Drude'' theory.
 
In analogy to (\ref{eq:rhoc.intra}),
the interband contribution of the $c$-axis conductivity is given by

\begin{eqnarray}
&&\sigma_{c}^{\rm inter}(\omega)=e^2{1\over \Omega}
\sum_{\bf k}v_\perp^2 \int_{-\infty}^{\infty}{dx\over 2\pi}~
{f(x)-f(x+\hbar\omega)\over \omega}\nonumber\\
&&~~\Biggl[{2\Gamma\over (x-\epsilon_+)^2+\Gamma^2}\
{2\Gamma\over (x+\hbar\omega-\epsilon_-)^2+\Gamma^2}\nonumber\\
&&~~~+ {2\Gamma\over (x-\epsilon_-)^2+\Gamma^2}\
{2\Gamma\over (x+\hbar\omega-\epsilon_+)^2+\Gamma^2} \Biggr],
\label{eq:rhoc.inter}
\end{eqnarray}
where we note that the second term dominates when $\omega\geq 0$
(because $\epsilon_+ \geq\epsilon_-$).
To calculate (\ref{eq:rhoc.inter}), the simplification 
(\ref{eq:sum.approxi.1}) used for the calculations of intraband part 
is no longer valid because now the two bands come in together and the
correlation and difference between these two bands are certainly important.
One therefore needs instead to replace the momentum
${\bf k}$-sum in (\ref{eq:rhoc.inter}) by the integral

\begin{eqnarray}
{1\over \Omega} \sum_{\bf k}&\rightarrow&
 \int_{-{\pi\over d}}^{\pi\over d}{dk_z\over 2\pi}
\int_{-{\infty}}^{\infty}{dk_x\over 2\pi}
\int_{-{\infty}}^{\infty}{dk_y\over 2\pi}.
\label{eq:sum.approxi.2}
\end{eqnarray}
To simplify the calculation, we will approximate
$\epsilon_+=\xi_1$ and  $\epsilon_-=\xi_2$ in 
the small-$t_\perp$ limit (see (\ref{eq:spectrum2})).
This simplification was not done in 
Ref.~\cite{AC96-1} where the ${\bf k}$-sum
is carried out directly by numerical integration over the first Brillouin zone.
Note that this means that they are effectively including the Van Hove 
singularity in the electronic density of states which we 
have taken as constant. The integration over $k_z$
can be carried out first to be 
$\int_{-\pi/d}^{\pi/d}{dk_z\over 2\pi}v_\perp^2 =t_\perp^2 d/2\hbar^2$.
Moreover, with 

\begin{eqnarray}
&&~~~\int_{-\infty}^\infty dk_x~ {1\over (A-\hbar^2 k_x^2/2m)^2
+\Gamma^2}\nonumber\\
&&={\sqrt{m}\pi\over \hbar}{1\over
\sqrt{A^2+\Gamma^2}\sqrt{\sqrt{A^2+\Gamma^2}-A}},
\label{eq:int4}
\end{eqnarray}
$\sigma_{c}^{\rm inter}$ in (\ref{eq:rhoc.inter}) can be reduced to
(changing the variable $dk_y\rightarrow d\epsilon_-=d\xi_2$)

\begin{eqnarray}
&&\sigma_{c}^{\rm inter}(\omega)=
{e^2t_\perp^2N(0)d\over \hbar^2}\sqrt{2}\pi\nonumber\\
&&\times\int_{-\infty}^{\infty}dx~ {f(x)-f(x+\hbar\omega)\over \omega}
\int_{-{\mu}}^{\infty}{d\epsilon_-\over \sqrt{\epsilon_-+\mu}}\nonumber\\
&&\Biggl[{\Gamma\over \Bigl[(x+\hbar\omega-\epsilon_-)^2+\Gamma^2\Bigr]
\sqrt{(x-\epsilon_- -\Delta)^2+\Gamma^2}}\nonumber\\
&&\times{\Gamma \over \sqrt{\sqrt{(x-\epsilon_- -\Delta)^2+\Gamma^2}
-(x-\epsilon_- -\Delta)}}
\label{eq:rhoc.inter.3} \\
&&+{\Gamma\over \Bigl[(x-\epsilon_-)^2+\Gamma^2\Bigr]
\sqrt{(x-\epsilon_-+\hbar\omega-\Delta)^2+\Gamma^2}}\nonumber\\
&&\times{\Gamma \over \sqrt{\sqrt{(x-\epsilon_-+\hbar\omega-\Delta)^2+\Gamma^2}
-(x-\epsilon_-+\hbar\omega-\Delta)}}\Biggr],\nonumber
\end{eqnarray}
with $N(0)\equiv m/2\pi\hbar^2$. A crucial point seen in 
(\ref{eq:rhoc.inter.3}) is that the chemical potential (or filling)
plays a role in the interband transition, as expected.

\begin{figure}
\vspace{-0.6cm}
\postscript{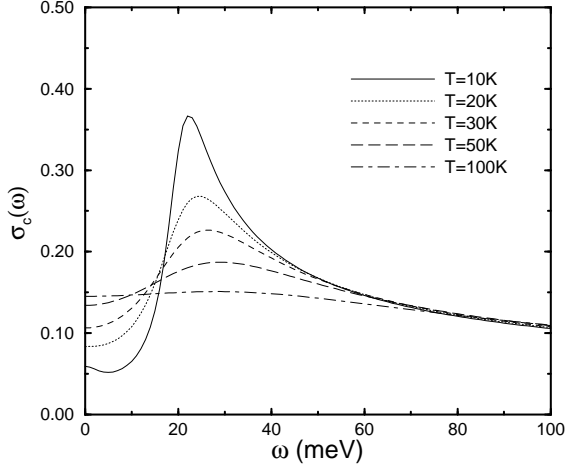}
\vspace{-0.5cm}
\caption{$c$-axis conductivity as a function of photon
energy for a plane-chain bilayer at different
temperatures with a small $t_\perp=2{\rm meV}$.
The minimum energy difference between the two bands for
a given ${\bf k}$ in the first 2D Brillouin zone is $\Delta=20{\rm meV}$.}
\label{fig1}
\end{figure}

\begin{figure}
\vspace{-0.6cm}
\postscript{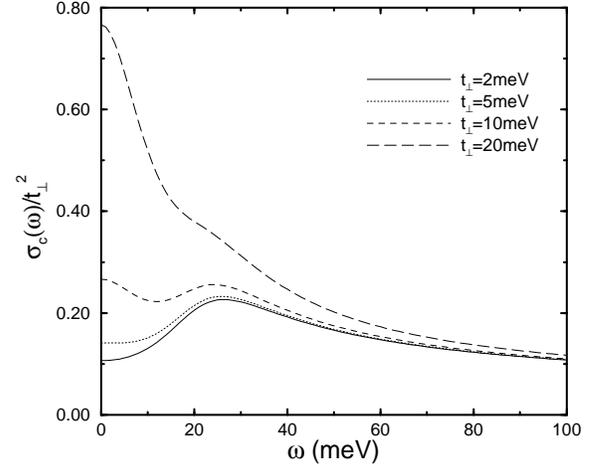}
\vspace{-0.5cm}
\caption{$c$-axis conductivity for a plane-chain bilayer for different
values of coupling $t_\perp$ at $T=30K$ with
minimum energy between the two bands equal to $\Delta=20{\rm meV}$.}
\label{fig2}
\end{figure}

Based on the intraband contribution given by the analytical
result (\ref{eq:Drude}) and the interband contribution given by the 
numerical calculations of (\ref{eq:rhoc.inter.3}), 
we have plotted the $c$-axis conductivity
in Fig.~\ref{fig1} for various temperatures at fixed $t_\perp=2{\rm meV}$
and in Fig.~\ref{fig2} for fixed temperature $T=30K$ at various $t_\perp$.
We have used the same parameters:
$\Delta=20{\rm meV}$, $\mu=500{\rm meV}$, and 
$\Gamma$ is taken to be linear in temperature with $\Gamma=20{\rm meV}$ at 
$T=100K$ throughout this paper unless specified.  
This is motivated by the fact that the inelastic scattering
rate is observed to be large and of 
the order of $T_c$ \cite{ZCP} at the superconducting 
critical temperature ($T_c$) and varying almost linearly
with temperature $T$. Also the evidence is that good samples show little
or no residual resistivity and are therefore in the pure limit 
(there is no residual elastic scattering).
One sees in Fig.~\ref{fig1} that
a pseudogap can develop at low temperatures for small $t_\perp$
(in which case the interband transitions dominate).
At higher temperatures, the pseudogap is invisible due to the
stronger inelastic damping effect.

While in this paper we have made simplifications
and approximations to the band structure, for example, 
we used the infinite-band approximation
(\ref{eq:sum.approxi.1}) with constant density of states evaluated at the
Fermi energy and taken it out of the integral,
this was not done in Ref.~\cite{AC96-1} where the full sum 
over the Brillouin zone which includes 
a Van Hove singularity in the electronic density of states
was performed numerically and
the effect of $t_\perp$ on the
bands (see Eq.~(\ref{eq:spectrum2})) was fully included.
We observe however that there are no qualitative differences between the
results of Fig.~\ref{fig1} and those of Fig.~2 in Ref.~\cite{AC96-1}.
The results of Ref.~\cite{AC96-1} do
show some structure at higher energy which reflects
special features of the plane-chain tight
binding bands used in that study but not in this work,
but these are small and we can safely proceed with a discussion of the DC 
resistivity using the simplified infinite band structure
of Eq.~(\ref{eq:simpleband}) which allows the required numerical
work to be greatly simplified. 

Fig.~\ref{fig2} shows that by increasing strength of $t_\perp$,
the pseudogap can disappear. 
For clarity of presentation, all intensities in Fig.~\ref{fig2}
have been divided by $t_\perp^2$. 
This leads to a nearly constant interband contribution
plus a intraband (Drude-like) contribution proportional to
$t_\perp^2$ which increases with increasing
values of $t_\perp$. In the case of larger $t_\perp$,
the intraband transition and hence the Drude-like behavior becomes dominant
(long dashed curve with $t_\perp=20{\rm meV}$) over the smaller interband 
contribution which now appears above $\sim 20{\rm meV}$.
We stress and it is clear from the figures that
the value of the real part of the conductivity
at higher energies is entirely due to
interband contributions since in our model the Drude contribution is
sharply peaked about $\omega=0$ on the energy scale of 150 to 200 meV.
 
\begin{figure}[h]
\vspace{-0.6cm}
\postscript{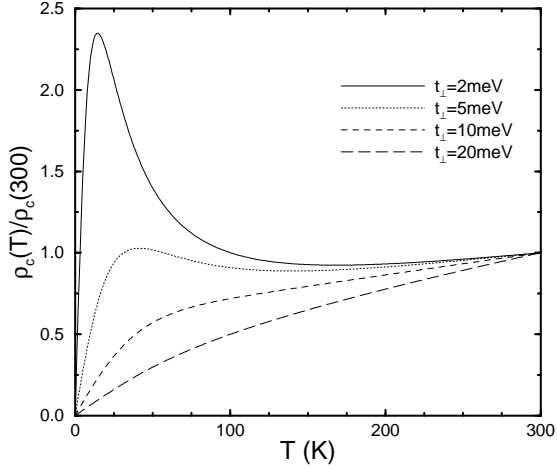}
\vspace{-0.5cm}
\caption{Temperature-dependent $c$-axis
DC resistivity for a plane-chain bilayer for different values of $t_\perp$.
The minimum energy between the two bands is $\Delta=20{\rm meV}$.}
\label{fig3}
\end{figure}

\begin{figure}[h]
\vspace{-0.6cm}
\postscript{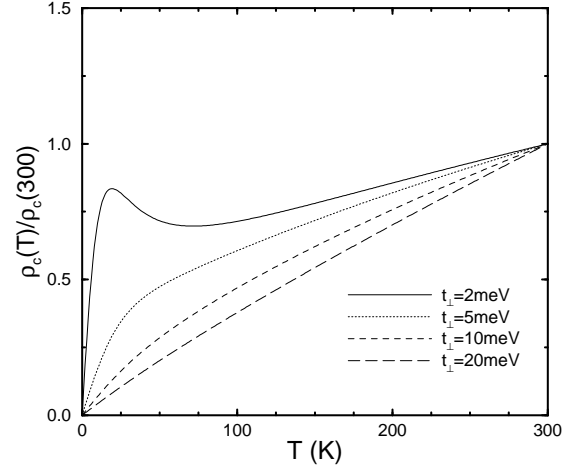}
\vspace{-0.5cm}
\caption{Similar plot as Fig.~\protect\ref{fig3}, but for
$\Delta=10{\rm meV}$.}
\label{fig4}
\end{figure}

We have also plotted in
Figs.~\ref{fig3} and \ref{fig4} the $c$-axis
DC resistivity for the plane-chain bilayer normalized to its value
at $300K$.
As shown in Fig.~\ref{fig3}, when $t_\perp$ is small
the $c$-axis DC resistivity shows a
characteristic upturn as the temperature is lowered
such as those observed in semiconductors. This characterizes
the importance of the interband transition.  
A ``pseudogap'' in the joint density of states (interband contribution)
suppresses the interlayer hopping and gives rise to the semiconductor-like
$\rho_c(T)$. When $t_\perp$ is increased, intraband transitions start to 
dominate and a more conventional response (linear in $T$) is obtained.
A similar plot as given in Fig.~\ref{fig3} given 
in Fig.~\ref{fig4} shows that
when the pseudogap is reduced (or relatively the $t_\perp$ is larger),
the upturn feature becomes much less dramatic.

To gain more insight into the relationship between
the $c$-axis conductivity and DC resistivity in a plane-chain bilayer, 
we derive in the following some analytic results under certain limits.
When temperature is not too high and hence $\Gamma$ is sufficiently small
(it is assumed to vary linearly with $T$),
one can approximate

\begin{equation}
{\Gamma\over (x-\epsilon_-)^2+\Gamma^2}\rightarrow \pi\delta(\epsilon_- -x)
\label{eq:replacement}
\end{equation}
in the second term of (\ref{eq:rhoc.inter.3}) which
one recalls dominates in the case of $\omega>0$ which is of interest. 
Moreover, due to the behavior of the Fermi distribution functions,
the integration over $x$ in (\ref{eq:rhoc.inter.3}) contributes mainly 
from $x=-\hbar\omega-k_BT$ to $x=k_BT$.
One thus can replace
$\int_{-{\mu}}^{\infty}d\epsilon_-$ by
$\int_{-\infty}^{\infty}d\epsilon_-$ assuming $\hbar\omega\ll \mu$.
Consequently, after the integration over $d\epsilon_-$,
the second term of (\ref{eq:rhoc.inter.3}) reduces to

\begin{eqnarray}
&&{e^2t_\perp^2N(0)d\over \hbar^2}\sqrt{2}\pi^2
\int_{-\infty}^{\infty}dx~
{f(x)-f(x+\hbar\omega)\over \omega}{1\over \sqrt{x+\mu}}\nonumber\\
&&{\Gamma\over \sqrt{(\hbar\omega-\Delta)^2+ \Gamma^2}
\sqrt{\sqrt{(\hbar\omega-\Delta)^2+\Gamma^2}
-(\hbar\omega-\Delta)}}.
\label{eq:rhoc.inter.5}
\end{eqnarray}
Eq.~(\ref{eq:rhoc.inter.5}) allows one
to discuss some analytic results of $\sigma_{c}^{\rm inter}$
(and hence $\sigma_{c}$).

For high frequencies ($\hbar\omega-\Delta\gg \Gamma, k_B T$),
one can approximate 
$\int_{-\infty}^{\infty}dx
\left[f(x)-f(x+\hbar\omega)\right]$ by
$\int_{-\hbar\omega}^{0}dx$,
and, as a result, the associated $x$-integration  can be performed easily
to be
 
\begin{eqnarray}
\int_{-\hbar\omega}^{0}dx~ {1\over \sqrt{x+\mu}} =
2(\sqrt{\mu}-\sqrt{\mu-\omega}).
\label{eq:rhoc.inter.7}
\end{eqnarray}
Next by expanding

\begin{eqnarray}
\sqrt{\sqrt{(\hbar\omega-\Delta)^2+\Gamma^2}-(\hbar\omega-\Delta)}
\simeq {\Gamma\over \sqrt{2(\hbar\omega-\Delta)}},
\label{eq:expand1}
\end{eqnarray}
one obtains following (\ref{eq:rhoc.inter.5})

\begin{eqnarray}
\sigma_{c}^{\rm inter}~\sim~ t_\perp^2{\sqrt{\mu}-\sqrt{\mu-\omega}
\over \omega\sqrt{\hbar\omega-\Delta}}, 
\label{eq:rhoc.inter.10}
\end{eqnarray}
which is {\em independent} of $\Gamma$ (or $T$)
and is proportional to $\omega^{-{1\over 2}}$ when
$\mu\gg\omega\gg \Delta$. 
 
For low frequencies ($\hbar\omega\ll \Delta$), one has
$[f(x)-f(x+\hbar\omega)]/\omega\rightarrow \hbar\delta(x)$,
and following (\ref{eq:rhoc.inter.5})

\begin{eqnarray}
&&~~\sigma_{c}^{\rm inter}(\omega)~\sim~t_\perp^2\mu^{-{1\over 2}} 
\label{eq:rhoc.inter.6} \\
&&\times{\Gamma\over
 \sqrt{(\hbar\omega-\Delta)^2+
\Gamma^2}\sqrt{\sqrt{(\hbar\omega-\Delta)^2+\Gamma^2}-(\hbar\omega-\Delta)}},
\nonumber
\end{eqnarray}
which is proportional to $\Gamma$ when $\Gamma\ll \Delta$ and
is proportional to
$\Gamma^{-{1\over 2}}$ when $\Gamma\gg \Delta$.
From the result (\ref{eq:rhoc.inter.6}) and (\ref{eq:Drude}) at $\omega=0$,
one can also get an expression for
the DC resistivity ($\rho_c(T)$) for the plane-chain bilayer,
which is then given by

\begin{eqnarray}
&&~~\rho_c(T)~ \sim ~t_\perp^{-2}\Gamma
\label{eq:dc.resis.pc} \\
&&\times\left[\left({t_\perp\over\Delta}\right)^2
+{\pi^2\over \sqrt{8}}{\Gamma^2\over \sqrt{\mu}\sqrt{\Delta^2+ \Gamma^2}
\sqrt{\sqrt{\Delta^2+\Gamma^2}+\Delta}}\right]^{-1}.
\nonumber
\end{eqnarray}
It is clear from (\ref{eq:dc.resis.pc}) that for
$t_\perp\ll \Delta$ the first term in the 
square bracket is small and becomes important only
when the second one is also small.
For $\Gamma$ small,

\begin{equation}
\rho_c(T)~ \sim
~t_\perp^{-2}\Gamma\left[\left({t_\perp\over\Delta}\right)^2
+{\pi^2\over 4}{\Gamma^2\over \sqrt{\mu\Delta^3}}\right]^{-1},
\label{eq:dc.resis.pc.1}
\end{equation}
the Drude contribution to the $c$-axis
resistivity $\rho_c(T)$ goes
like $\Gamma$ while the interband contribution on its own
has the inverse dependence
and goes like $\Gamma^{-1}$. This is the physics underlying
the upturn in the resistivity of Figs.~\ref{fig3} and
\ref{fig4} as $T$ is lowered and is also the reason the pseudogap
forms as seen in Fig.~\ref{fig1}.
The two effects are intimately connected in our theory.

\begin{figure}[h]
\vspace{-0.6cm}
\postscript{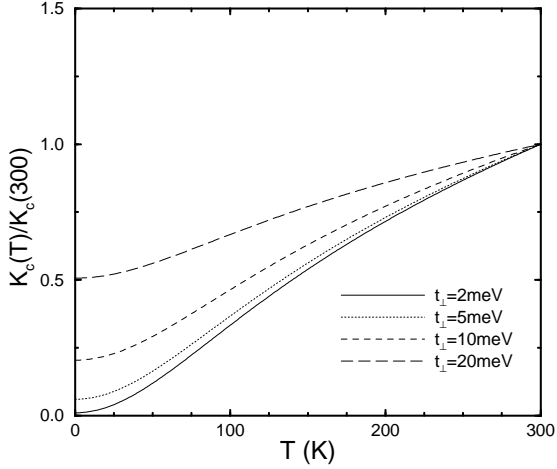}
\vspace{-0.5cm}
\caption{Temperature-dependent $c$-axis thermal conductivity
for a plane-chain bilayer for different values of $t_\perp$.}
\label{fig5}
\end{figure}
 
\section{Thermal conductivity}

In analogy to (\ref{eq:conductivity}) for the $c$-axis optical
conductivity, one can derive the $c$-axis thermal conductivity
in terms of the Kubo formalism  \cite{AG,Scalapino}
 
\begin{eqnarray}
&&K_{c}(T)={1\over T\Omega}
\sum_{\bf k}\int_{-\infty}^{\infty}{dx\over 2\pi}~x^2
{\partial f(x)\over \partial x} \nonumber\\
&&\times {\rm Tr} [A({\bf k},x)\gamma_z({\bf k,k})
A({\bf k},x)\gamma_z({\bf k,k})],
\label{eq:thermoconductivity}
\end{eqnarray}
calculated at $\omega=0$. Here $f(x)$ is the Fermi distribution function,
$A({\bf k},x)$ is the spectral function matrix, and
$\gamma_z({\bf k,k})$ is exactly the vertex function matrix for calculating
the current-current correlation function given in (\ref{eq:gammaz.def}).
As we have done in (\ref{eq:rhoc}) for the optical conductivity,
due to the trace operator, one can work in a frame in which the
Green's function is diagonal.
This will lead to the fact that
the $c$-axis thermal conductivity
can be separated into intraband and interband contributions.
However, this is not needed since we can establish that
the $c$-axis thermal conductivity obeys the Wiedemann-Franz law.
Due to the fact that the main contribution to integration
over $x$ for both (\ref{eq:thermoconductivity}) and (\ref{eq:conductivity})
at $\omega=0$ comes from the small region around $x=0$,
the $x$-dependence inside the trace 
for both (\ref{eq:thermoconductivity}) and (\ref{eq:conductivity})
is removed. One easily finds by comparison that the ratio of 
the $c$-axis thermal conductivity to the
$c$-axis DC electrical conductivity is
 
\begin{eqnarray}
&&~~{K_c\over \sigma_c(\omega=0)}={1\over T}\int_{-\infty}^{\infty}dx~x^2
{\partial f(x)\over \partial x}\left/
e^2\int_{-\infty}^{\infty}dx~
{\partial f(x)\over \partial x}\right.\nonumber\\
&&=\left({k_B\over e}\right)^2 T\int_{-\infty}^{\infty}dy~2y^2
{\rm sech}^2 y= {\pi^2\over 3}\left({k_B\over e}\right)^2 T.
\label{eq:WF.law}
\end{eqnarray}
The above result is simply the Wiedemann-Franz law.
 
Using (\ref{eq:dc.resis.pc}) with (\ref{eq:WF.law}), 
one obtains (recall $\Gamma\sim T$)
 
\begin{eqnarray}
&&~~~K_c(T)~\sim ~t_\perp^{2}
\label{eq:Kc.final}\\
&&\times\left[\left({t_\perp\over\Delta}\right)^2
+{\pi^2\over \sqrt{8}}{\Gamma^2\over \sqrt{\mu}\sqrt{\Delta^2+ \Gamma^2}
\sqrt{\sqrt{\Delta^2+\Gamma^2}+\Delta}}\right], \nonumber
\end{eqnarray}
and for small $\Gamma$ (using (\ref{eq:dc.resis.pc.1}))

\begin{equation}
K_c(T)~\sim ~t_\perp^{2} \left[\left({t_\perp\over\Delta}\right)^2
+{\pi^2\over 4}{\Gamma^2\over \sqrt{\mu\Delta^3}}\right].
\label{eq:Kc.final.1}
\end{equation}
We have plotted $K_c(T)$ normalized to its value at $300K$
using (\ref{eq:Kc.final}) in Fig.~\ref{fig5}. 
For small $t_\perp$, the significant depression of 
$K_c(T)$ at low temperature characterizes the importance of the interband 
transition in which case one will also see the depression of optical
conductivity and the upturn of the DC resistivity
associated with pseudogaps. 

\section{Plane-Plane Bilayer}

Even simpler analytic results, which shed further light on the
physics of our model can be obtained for 
a plane-plane bilayer in which the two uncoupled bands are given by
(\ref{eq:simpleband}) with $\xi_2$ now replaced by
$\xi_2={\hbar^2 \over 2m}(k_x^2+k_y^2)-\mu$. In this case, one can 
also have analytical results
for the interband part of the $c$-axis conductivity
 
\begin{eqnarray}
&&\sigma_{c}^{\rm inter}(\omega)=
{e^2t_\perp^2N(0)d\over \hbar}
\left[{2\Gamma\over (\hbar\omega+\Delta)^2+4\Gamma^2}\right.\nonumber\\
&&~~~~~~~~~+\left. {2\Gamma\over (\hbar\omega-\Delta)^2+4\Gamma^2}\right].
\label{eq:Drude.inter}
\end{eqnarray}
The result (\ref{eq:Drude.inter}) for $\sigma_{c}^{\rm inter}$
for the plane-plane bilayer leads to a 
peak feature at $\omega=\Delta$ but is otherwise similar to the results
given by (\ref{eq:rhoc.inter.5}) (or more generally (\ref{eq:rhoc.inter.3}))
for a plane-chain bilayer. The simple analytic 
expression (\ref{eq:Drude.inter}) thus 
helps gain physical insight into the $c$-axis transport properties.

\subsection{\bf Conductivity}

The total $c$-axis conductivity given by the sum of
(\ref{eq:Drude}) and (\ref{eq:Drude.inter}) will be

\begin{eqnarray}
&&\sigma_{c}(\omega)=\sigma_{c}^{\rm intra}(\omega)
+\sigma_{c}^{\rm inter}(\omega)=
{e^2t_\perp^2N(0)d\over \hbar} \nonumber\\
&&~~~~\times\left[\left({2t_\perp\over\Delta}\right)^2
{4\Gamma\over (\hbar\omega)^2+4\Gamma^2}\right.
\label{eq:Drude.total}\\
&&\left.+\left({2\Gamma\over (\hbar\omega+\Delta)^2+4\Gamma^2}+
{2\Gamma\over (\hbar\omega-\Delta)^2+4\Gamma^2}\right)\right].
\nonumber
\end{eqnarray}
Some features are: 
 
\begin{itemize}
\item $\sigma_{c}^{\rm intra} \propto t_\perp^4$,
$\sigma_{c}^{\rm inter}\propto t_\perp^2$, and the
relative intensity $\sigma_{c}^{\rm intra}/
\sigma_{c}^{\rm inter}\propto t_\perp^2/\Delta^2$.
\item When $\omega\rightarrow 0$ and $T$ (or $\Gamma$) is small,
$\sigma_{c}^{\rm intra}\sim \Gamma^{-1}\sim T^{-1}$ (Drude like),
while $\sigma_{c}^{\rm inter}\sim \Gamma\sim T$ (non-Drude like).
\item The high-frequency tail of $\sigma_{c}(\omega)$
goes as $\omega^{-2}$ and decays very fast.
\item The peak at $\omega=\Delta$ corresponds to the pseudogap
and by changing the ratio of $t_\perp$ to $\Delta$, 
the pseudogap can disappear.
\end{itemize}
 
\subsection{\bf Frequency Sum Rule}
 
Because of
 
\begin{eqnarray}
&&\int_{0}^\infty d\omega {4\Gamma\over (\hbar\omega)^2+4\Gamma^2}
=\int_{0}^\infty d\omega\left[{2\Gamma\over (\hbar\omega+\Delta)^2+4\Gamma^2}
\right.\nonumber\\
&&\left.~~~~~~~~~~+
{2\Gamma\over (\hbar\omega-\Delta)^2+4\Gamma^2}\right]={\pi\over \hbar},
\label{eq:int3}
\end{eqnarray}
one can easily obtain
 
\begin{equation}
\int_{0}^\infty d\omega~\sigma_{c}(\omega)=
{e^2t_\perp^2N(0)d\pi\over \hbar^2}
\left[1+\left({2t_\perp\over\Delta}\right)^2
\right],
\label{eq:sum.rule}
\end{equation}
which states the $\omega$-sum rule for the optical conductivity.
 
\subsection{\bf DC Resistivity}
 
Using the result (\ref{eq:Drude.total}) in the limit of $\omega\rightarrow 0$,
we obtain the $c$-axis DC resistivity
 
\begin{equation}
\rho_c(T)=[\sigma_{c}(\omega\rightarrow 0)]^{-1}
\sim
t_\perp^{-2}\Gamma\left[\left({t_\perp\over\Delta}\right)^2
+{\Gamma^2\over \Delta^2+4\Gamma^2}\right]^{-1}.
\label{eq:dc.resis}
\end{equation}
The semiconductor-like upturn behavior at low temperature is obtained
when the second term (interband transition) is dominant.
 
\subsection{\bf Thermal conductivity}

Based on Wiedemann-Franz law
and using (\ref{eq:Drude.total}) at $\omega=0$
or directly from (\ref{eq:dc.resis}), one obtains 
the $c$-axis thermal conductivity for the plane-plane bilayer

\begin{equation}
K_c(T)~\sim ~t_\perp^{2}\left[\left({t_\perp\over\Delta}\right)^2
+{\Gamma^2\over \Delta^2+4\Gamma^2}\right].
\label{eq:Kc.final.pp}
\end{equation}
Once again, $K_c(T)$ will show strong depression at low temperature
when the interband transition dominates.

\section{Case of Tight-Binding Bands}

\begin{figure}
\begin{center}
\leavevmode
\epsfxsize 0.9\columnwidth
\epsffile{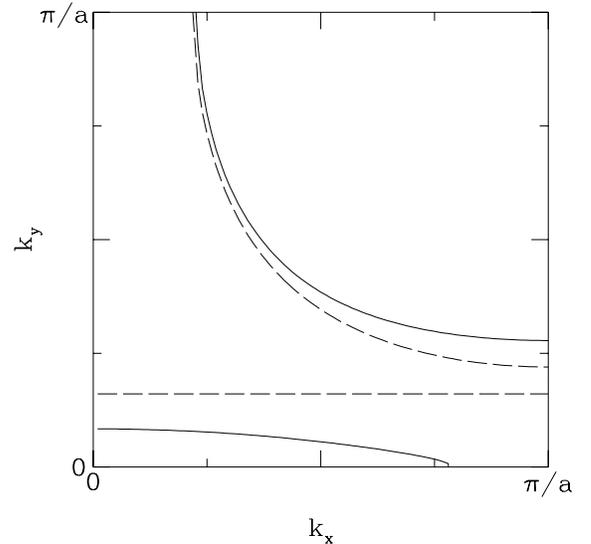}
\caption{The structure of the tight-binding bands given in
Eq.~(\protect\ref{eq:band}). The parameters are
$\{t_1,t_2,\mu_1,\mu_2,t_\perp\}$ = $\{70,100,-65,-175,20\} \mbox{ meV}$ and
$B = 0.45$.}
\label{fig6}
\end{center}
\end{figure}

So far we have used the simplified band
structures (\ref{eq:simpleband}) and have tried to
get simple analytic final results.
These have required some approximations which we now wish
to remove. We start with tight-binding bands with momentum ${\bf k}$
confined to the first Brillouin zone of the copper-oxide
plane. We take

\begin{mathletters}
\label{eq:band}
\begin{eqnarray}
\xi_1=-2t_1[\cos(k_x a)+\cos(k_y a)\nonumber\\
-2B\cos(k_x a)\cos(k_y a)]-\mu_1
\label{eq:band1}
\end{eqnarray}
and

\begin{eqnarray}
\xi_2=-2t_2\cos(k_y a)-\mu_2.
\label{eq:band2}
\end{eqnarray}
\end{mathletters}
For a model with $\{t_1,t_2,\mu_1,\mu_2,t_\perp\}
= \{70,100,-65$, $-175, 20\}$ in meV and $B=0.45$,
the corresponding Fermi surface contours are
shown in Fig.~\ref{fig6}.
The dashed line are the unperturbed Fermi
surface which have been chosen so that they do not
cross. They are also the contours for the hybridized case with
finite $t_\perp$ and $k_z={\pi\over d}$ 
in which case $t(k_z)$ in Eq.~(\ref{eq:t})
is zero. The solid lines are the contours for $k_z=0$ where
$t(k_z)$ has its maximum value in (\ref{eq:t}).
The area between dashed and solid curves gives the dispersion
in the $z$-direction of the Fermi surface
in the three-dimensional Brillouin zone.
For this model band structure, we have evaluated (\ref{eq:rhoc}) 
without further approximations (using a three-dimensional numerical routine 
over the Brillouin zone as in our previous paper \cite{AC96-1}) and
found results (see Fig.~2 in Ref.~\cite{AC96-1})
for the frequency dependence of the $c$-axis
conductivity in qualitative agreement with our Fig.~\ref{fig1}.

\begin{figure}
\begin{center}
\leavevmode
\epsfxsize 1.0\columnwidth
\epsffile{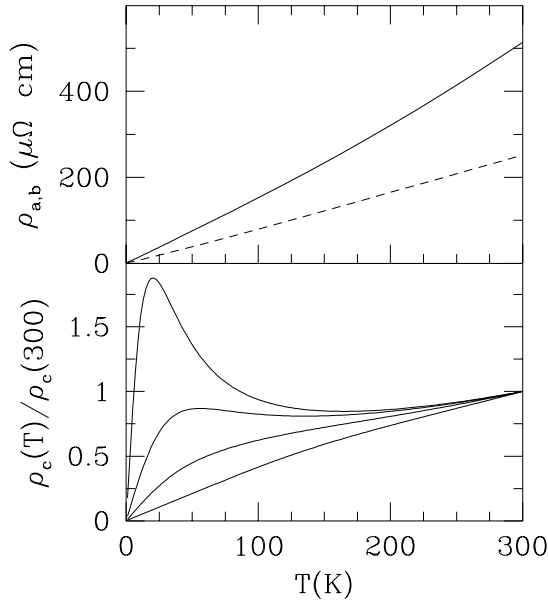}
\caption{Resistivity in the $a$ (top frame, solid line), $b$
(top frame, dashed line) and $c$ (bottom frame) directions using the
tight-binding bands. The $a$ and $b$-axis resistivity are for
$t_\perp=2{\rm meV}$ and the $c$-axis
resistivities are for $t_\perp = 2$,  5, 10, and 20 meV
from top to bottom.}
\label{fig7}
\end{center}
\end{figure}
 
In Fig.~\ref{fig7}, we show results for the DC resistivity.
The top frame refers to the $a$- and $b$-directions,
while the lower frame is for the $c$-axis resistivity $\rho_c(T)$
normalized to its value at $300K$. Four values of $t_\perp$ have been
used for $\rho_c(T)$ namely $t_\perp=2, 5, 10$ and 20 meV from
the top to bottom. On comparing this frame with our previous 
Fig.~\ref{fig3}, it is clear that there is no qualitative differences
between the two models and use of the more realistic band structures has
made only minor difference in our results for the DC resistivity.
In the top frame of Fig.~\ref{fig7} ($t_\perp=2{\rm meV}$), the solid line
is for the $a$-direction and the dashed line for the $b$-direction 
which is along the chain. 
While interband contributions have been included in the
calculations with different values of $t_\perp$, 
this was found to make only small differences on the scale
shown and so only a 
single curve is shown. For even higher values of $t_\perp$, larger 
differences should start showing up. In the case 
considered, the $a$-$b$ plane resistivity
$\rho_{ab}(T)$ is nearly linear in $T$ but not quite. 
We note that our phenomenological assumption that the 
in-plane scattering rate $\Gamma(T)$ is linear in
$T$ leads directly to a linear dependence of $\rho_{ab}(T)$ vs. $T$ in
an infinite band model with
constant density of states. The dashed line
in the top frame of Fig.~\ref{fig7} is seen to be close to linear but
this is not as true for the solid curve (along the $a$-direction) which 
shows two temperature scale -- the second presumably related to the
Van Hove singularity in the two-dimensional 
tight-binding band structures used here.
There is also a small interband contribution 
further complicating the situation.

\section{Conclusions}

In this paper we have derived approximate expressions for
the frequency dependence of the $c$-axis conductivity 
for a bilayer cuprate based on simple
free electron bands. The simplified 
results reproduced well our previous numerical
results based on tight-binding bands \cite{AC96-1}. 
The advantage of the present
approach is that analytic results can be obtained in several important 
cases. Besides the AC infrared conductivity, we also
give results for the DC resistivity,
the optical conductivity sum rule and the thermal conductivity
via the Wiedemann-Franz law. In the
case of the $c$-axis resistivity,  results of full numerical
calculations based on tight-binding 
bands not presented before
are also given and compared with the free electron case.
In the discussion emphasis is placed on the role of
the pseudogap which is identified as the {\em minimum} energy
for interband transition possible in a given band structure.

\acknowledgments
We thank Dwayne Branch for useful discussion.
This work was supported by Natural Sciences and Engineering Research Council
(NSERC) of Canada and Canadian Institute for Advanced Research (CIAR).
One of the authors (W.A.) was supported in part
by the Midwest Superconductivity Consortium through D.O.E. 
grant \# DE-FG-02-90ER45427.

\end{document}